\documentclass{ws-procs9x6}

\begin{document}

\title{Structure of Neutron-Rich Nuclei in the $^{132}$S\lowercase{n} Region }

\author{A. COVELLO, L. CORAGGIO, A. GARGANO, N. ITACO} 

\address{Dipartimento di Scienze Fisiche, Universit\`a di Napoli Federico II, \\
and Istituto Nazionale di Fisica Nucleare, \\
Complesso Universitario di Monte S. Angelo, Via Cintia, I-80126 Napoli, Italy \\ 
E-mail: covello@na.infn.it}


\maketitle

\abstracts {We report on a study of neutron-rich nuclei around $^{132}$Sn in terms of the shell model employing a realistic effective interaction derived from the CD-Bonn nucleon-nucleon potential. We present results for some Sb and Te isotopes. Comparison shows that our results are in very good agreement with the available experimental data supporting confidence in the predictions of our calculations. This may stimulate experimental efforts to gain more information on these nuclei lying well away from the valley of stability. }

\section{Introduction}

The study of neutron-rich nuclei in the $^{132}$Sn region is a subject of great interest. This is related to the fact that $^{132}$Sn is a very good doubly magic nucleus whose neighbors provide the opportunity for testing the basic ingredients of shell-model calculations, especially the matrix elements of the effective interaction, well away from the valley of stability.

From the experimental point of view, it is a very hard task to obtain information on these nuclei. In the last few years, however, substantial progress has been made to access the limits of nuclear stability, which has paved the way to spectroscopic studies in the 
$^{132}$Sn region. A summary of recent experimental efforts in this area including references through 2000 is given in Ref. 1.

Motivated by these experimental achievements, in recent years we have 
studied\cite{cov01,cor132} several nuclei around $^{132}$Sn  in terms of the shell model employing realistic effective interactions derived from modern nucleon-nucleon ($NN$) potentials. 

The main aim of this paper is to report on some selected results of our current work in this region, which have been obtained starting from the CD-Bonn free $NN$ potential.\cite{machl01} In particular, we shall consider two odd-odd antimony isotopes, $^{130,132}$Sb,
and three even-odd tellurium isotopes, $^{133,135,137}$Te.

As regards the former, we shall focus attention on the proton particle-neutron hole multiplets which play a special role for the understanding of the neutron-proton interaction around closed shells. More than thirty years ago the study of these multiplets in the Pb region was the subject of great experimental and theoretical interest.\cite{ersk64,alf68,kuo68,moine69} In this region, through pick-up and stripping reactions, several particle-hole multiplets were 
identified\cite{ersk64,alf68} in $^{208}$Bi. In this context, it is worth mentioning that a very good agreement with experiment was obtained in Ref. 7 using
particle-hole matrix elements deduced from the Hamada-Johnston potential.\cite{hama62}

Despite these early achievements in the study of the neutron-proton
interaction in the vicinity of doubly magic $^{208}$Pb, little work has been done ever since. By considering the new data which are becoming available in the $^{132}$Sn region, it is high time to revive theoretical interest in  this subject and perform shell-model calculations making us of a modern $NN$ potential and improved many-body methods for deriving the effective interaction.

In Sec. 2 we give a bare outline of the theoretical framework in which our realistic shell-model calculations have been performed. In Sec. 3 we present and discuss our results comparing them with the available experimental data. Sec. 4 presents some concluding remarks.
 
\section{Theoretical framework}

We assume that $^{132}$Sn is a closed core and let the valence protons and   
neutron holes occupy the five single-particle levels $0g_{7/2}$, $1d_{5/2}$, $1d_{3/2}$, $2s_{1/2}$,
and $0h_{11/2}$ of the 50-82 shell. Similarly, for the valence neutrons in $^{135,137}$Te the model space includes
all the six single-particle levels 
$0h_{9/2}$, $1f_{7/2}$, $1f_{5/2}$, $2p_{3/2}$, $2p_{1/2}$, and $0i_{13/2}$ 
of the 82-126  shell. The single-proton and single-hole energies have been taken from the
experimental spectra\cite{sanc99,foge84a,foge84b} of $^{133}$Sb  and $^{131}$Sn, respectively. The only exception is the proton $\epsilon_{s_{1/2}}$ which was taken from Ref. 13, since the corresponding single-particle level is still missing in $^{133}$Sb.
As regards the single-neutron energies, they have been taken from the
experimental spectrum\cite{hoff96} of $^{133}$Sn , except that relative to the
$i_{13/2}$ level which has not been observed. The latter has been taken from Ref. 15.

As already mentioned in  the Introduction, in our shell-model calculations we have made use of a realistic effective interaction derived from the CD-Bonn free nucleon-nucleon potential.\cite{machl01} This high-quality $NN$ potential, which is based upon meson exchange, fits very accurately ($\chi^2$/datum $\approx 1$) the world $NN$ data below 350 MeV available in the year 2000.

The shell-model effective interaction $V_{\rm eff}$ is defined, as usual, in the following way. In principle, one should solve a nuclear many-body Schr\"odinger equation of the form 
\begin{equation}
H\Psi_i=E_i\Psi_i 
\end{equation}
with $H=T+V_{NN}$, where $T$ denotes the kinetic energy. This full-space many-body problem is reduced to a smaller model-space problem of the form
\vspace{-.1cm}
\begin{equation}
PH_{\rm eff}P \Psi_i= P(H_{0}+V_{\rm eff})P \Psi_i=E_iP \Psi_i .
\end{equation}
\noindent Here $H_0=T+U$	 is the unperturbed Hamiltonian, $U$ being an auxiliary potential introduced to define a convenient single-particle basis, and $P$ denotes the projection operator onto the chosen model space, which generally consists of a major shell above the doubly closed core.

A main difficulty one is confronted with in the derivation of $V_{\rm eff}$ from a modern $NN$ potential, such as CD-Bonn, is the existence of a strong repulsive core which prevents its direct use in nuclear structure calculations. This difficulty is usually overcome by resorting to the time-honored Brueckner $G$-matrix method. Here, we have made use of a new approach\cite{bogn02} which provides an advantageous alternative to the use of the above method. It consists in constructing a low-momentum $NN$ potential, $V_{low-k}$, that preserves the physics of the original potential $V_{NN}$ up to a certain cut-off momentum $\Lambda$. In particular, the scattering phase shifts and deuteron binding energy calculated by $V_{NN}$ are reproduced by $V_{low-k}$. The latter is a smooth potential that can be used directly as input for the calculation of shell-model effective interactions. A detailed description of our derivation of $V_{low-k}$ can be found in Ref. 16,
where a criterion for the choice of the cut-off parameter $\Lambda$ is also given. We have used here the value $\Lambda=2.1$ fm$^{-1}$. 

Once the $V_{low-k}$ is obtained, the calculation of the matrix elements of the particle-particle and hole-hole interaction  is carried out within the framework of a folded-diagram method, as described, for instance, in Refs. 17 and 18.
It should be pointed out that for $^{130,132}$Sb and $^{133}$Te the neutron-proton effective interaction has been explicitly derived in the particle-hole formalism. 
A description of the derivation of the particle-hole effective interaction is given in Ref. 3.

\section{Results}

\begin{figure}[t]
\begin{center}
\epsfxsize=6cm   
\epsfbox{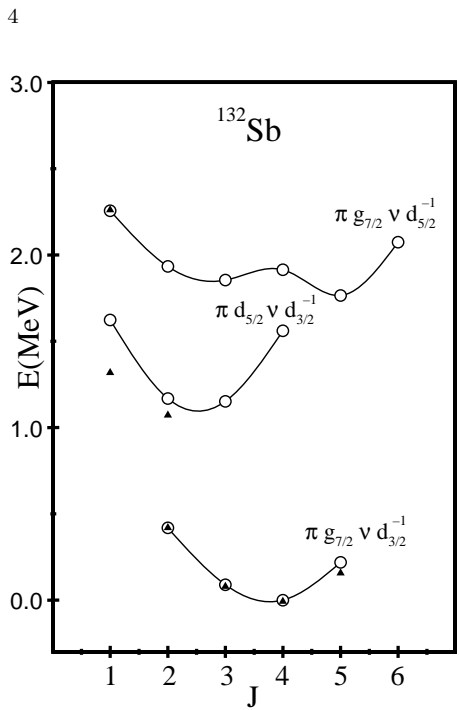}
\caption{Proton particle-neutron hole multiplets in $^{132}$Sb. The theoretical reults are represented by open circles while the experimental data by solid triangles. The lines are drawn to connect the points.} 
\end{center}
\vspace{-.2cm}
\end{figure}

We report here some selected results of our study of neutron-rich nuclei in the 
$^{132}$Sn region. All calculations have been performed using the OXBASH shell-model code.\cite{OXBASH}

We start by considering the Sb isotopes.  Some calculated multiplets for $^{132}$Sb are reported in Fig. 1 and compared with the existing 
experimental \hbox{data.\cite{ston89,mach95}} We see that the calculated energies are in very good agreement with the observed ones. In fact, the discrepancies are all in the order of tens of keV, except for the $1^+$ state of
the $\pi d_{5/2}$ $\nu d^{-1}_{3/2}$ multiplet, which lies 300 keV above the experimental counterpart. Some other calculated multiplets having the neutron hole in the $h_{11/2}$ level are reported in Ref. 3.

\begin{figure}[htb]
\begin{center}
\epsfxsize=6cm   
\epsfbox{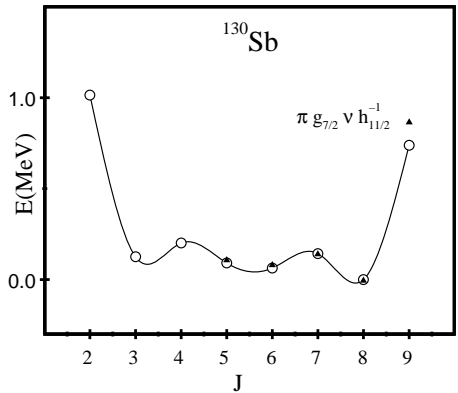}
\caption{Same as Fig. 1 \, but for the $\pi g_{7/2}$ $\nu h^{-1}_{11/2}$ multiplet in $^{130}$Sb.} 
\end{center}
\vspace{-.4cm}
\end{figure}

As regards $^{130}$Sb,  we report in Fig. 2 the $\pi g_{7/2}$ $\nu h^{-1}_{11/2}$ multiplet, for which several members have been experimentally 
identified.\cite{walte94,gene02} We see that the agreement between experiment and theory is remarkably good. It should be noted that both the experimental and calculated energies are relative to the $8^-$ state, which has been experimentally observed to be the ground state. Our calculations predict\cite{cor132} for the ground state $J^\pi=4^+$, the $8^-$ state lying at 92 keV excitation energy.

\begin{figure}[hbt]
\begin{center}
\epsfxsize=7cm   
\epsfbox{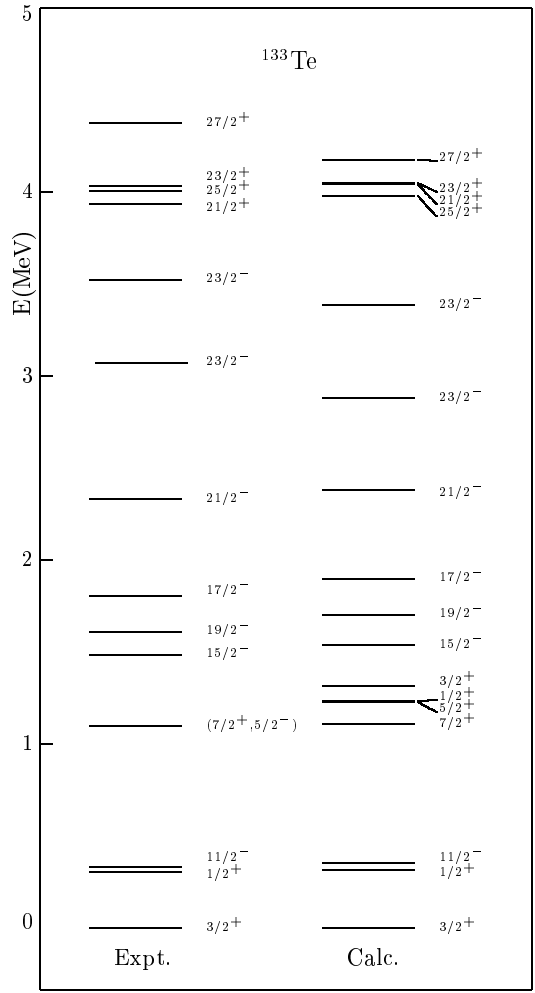}
\caption{Experimental and calculated spectra of $^{133}$Te.} 
\end{center}
\vspace{-.2cm}
\end{figure}

It is evident from Figs. 1 and 2 that a main feature of the calculated multiplets 
is that the states with minimum and maximum $J$ have the highest excitation energy and are well separated from the other states, for which the splitting is relatively small. This pattern is in agreement with the experimental one for the $\pi g_{7/2}$ $\nu d^{-1}_{3/2}$ multiplet in $^{132}$Sb and the experimental data available for the other multiplets also go in the same direction. It should be pointed out that this behavior  is quite similar to that exhibited\cite{kuo68,moine69} by the multiplets in the heavier particle-hole nucleus $^{208}$Bi. Also, it is worth noting that in all of our calculated multiplets (including those reported in Ref. 3), the state of spin ($j_\pi + j_\nu -1$) is the lowest, in agreement with the early predictions of the Brennan-Bernstein coupling rule.\cite{bren60}

We turn now to the tellurium isotopes, $^{133,135,137}$Te. These three nuclei have been the subject of recent experimental studies,\cite{bhat01,hwan02,forn01,urba00} where high-spin states of a particularly simple structure were identified.

The calculated spectrum of $^{133}$Te  is compared with the experimental one 
in Fig. 3 . We include all the observed\cite{nndc} and calculated states up to 1.3 MeV. In the higher energy region, where several states with unknown or ambiguous spin and parity have been observed, we only include the high-spin states reported in Refs. 25 and 26. As regards the calculated spectrum, we report the states which can be safely associated with the experimental ones. As for the agreement between theory and experiment, we see that all the observed excited levels are very well reproduced by the theory, the discrepancies being well below 100 keV for most of them. A detailed discussion of the structure of the calculated states can be found in Ref. 3.

\begin{figure} [htb]
\begin{center}
\epsfxsize=7cm   
\epsfbox{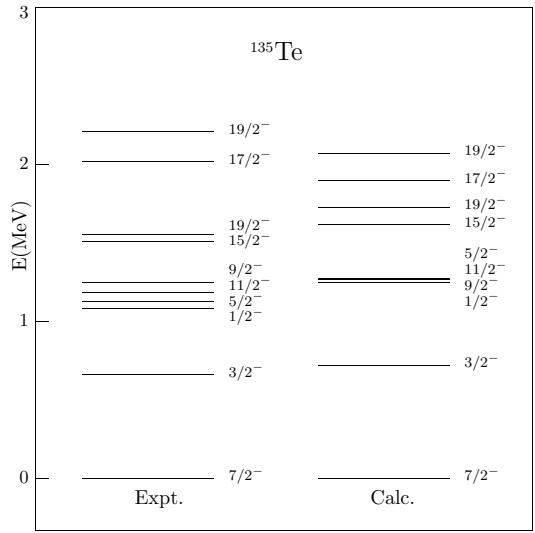}
\caption{Experimental and calculated spectra of $^{135}$Te.} 
\end{center}
\end{figure}

\begin{figure} [htb]
\begin{center}
\epsfxsize=7cm   
\epsfbox{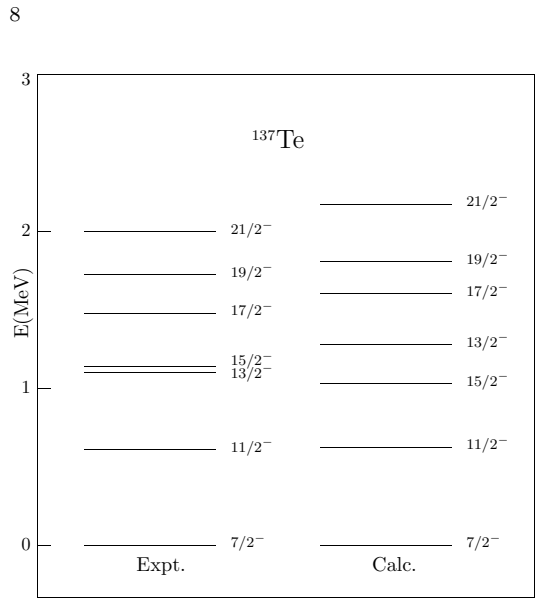}
\caption{Experimental and calculated spectra of $^{137}$Te.} 
\end{center}
\end{figure}

The theoretical spectrum of $^{135}$Te  is compared with the experimental one 
in Fig. 4, where all the observed\cite{nndc} and calculated states up to 1.3 MeV are included. In the higher energy region up to about 2.2 MeV we compare the four observed high-spin states\cite{forn01} with those predicted by the theory. 
As regards $^{137}$Te, in Fig. 5 we compare the experimental levels\cite{urba00} below 2 MeV with  those predicted by our calculations.

From Figs. 4 and 5 we see that the calculated spectra are in quite good agreement 
with the experimental ones, the largest discrepancy being about 170 keV in both cases.

\section{Concluding remarks}

We have presented here some results of a shell-model study of neutron-rich nuclei close to doubly magic $^{132}$Sn, focusing attention on the two antimony isotopes  $^{130,132}$Sb and the three tellurium isotopes $^{133,135,137}$Te, for which new relevant information has been obtained in recent experimental studies.
In our calculations we have employed a realistic effective interaction derived from the CD-Bonn $NN$ potential. This has been done within the framework of a new 
approach\cite{bogn02} to shell-model effective interactions which provides an advantageous alternative to the usual Brueckner $G$-matrix method.
We have shown that our results are in very good agreement with the  experimental data for all the nuclei considered. It should be stressed that our calculations are free from adjustable parameters.

On the above grounds, we may conclude with the following remarks.

(i) Effective interactions derived from modern $NN$ potentials are able to describe with quantitative accuracy the spectroscopic properties of nuclei in the $^{132}$Sn region far from stability. This gives confidence in their predictive power.

(ii) On the experimental side, it is of utmost importance to gain more information on neutron-rich nuclei in this  region. This is certainly a very exciting physics to be done with radioactive ion beams. 
 
\section*{Acknowledgments}
This work was supported in part by the Italian Ministero dell'Istruzione, dell'Universit\`a e della Ricerca (MIUR).

\end{document}